Synchronized excitation of magnetization dynamics via spin waves in Bi-YIG thin film by slot line waveguide

Tetsunori Koda[1], Sho Muroga[2] and Yasushi Endo[3,4,5]

[1] General Education Division, National Institute of Technology, Oshima College, Suo-Oshima 742-2193, Japan

[2] Department of Mathematical Science and Electrical-Electronic-Computer Engineering、Graduate School of Engineering Science, Akita University, Akita 010-8502, Japan

[3] Department of Electrical Engineering、Graduate School of Engineering, Tohoku University, Sendai 980-8579, Japan

[4] Center for Spintronics Research Network, Tohoku University, Sendai 980-8577, JAPAN

[5] Center for Science and Innovation in Spintronics (CSIS), Organization for Advanced Studies, Tohoku University, Sendai 980-8577, JAPAN

**ABSTRACT**

We have studied magnetization dynamics of single Bi-YIG thin films by means of the high frequency power response induced by a slot line waveguide. Multiple absorption peaks that correspond to excitement states in magnetization dynamics appeared without the ferromagnetic resonance (FMR) condition. The peaks were strongly influenced by a waveguide line width and a distance between the lines. Micromagnetics simulation reveals that each line induces a local magnetization dynamics oscillation and generates spin waves. The spin wave that propagates from one of the lines interferences with the other side of local magnetization dynamics oscillation around the other line, resulting in an amplification of the oscillation when they are in synchronization with each other. This amplification occurs at both sides of the lines by the interference. Thus, the possible mechanism of the excitation in the magnetization dynamics oscillation is the synchronization of mutual magnetization dynamics oscillation via spin waves. This technique resonantly excites the local magnetization dynamics without the FMR condition, which is applicable as a highly coherent spin waves source.

When a magnetic field is applied to a ferromagnetic material, the magnetization precesses around the axis of final state direction. The precession motion of magnetization, magnetization dynamics, is a basic phenomenon of magnetism. Especially, near the ferromagnetic resonance (FMR) conditions, the precession motion is effectively excited. Since the precession frequency is of the order of GHz, this phenomenon is a strong interest to both fundamental physics and an engineering points of view. For example, a spin torque nano oscillator shows promising properties for microwave emission [1-3] and reservoir computing [4-5].

At near FMR condition, a response of magnetization precession motion shows strong nonlinearity with respect to the change of an applied magnetic field applicable for highly sensitive magnetic sensors. We have reported that the change in magnetization dynamics was sensitive enough to detect a small change in magnetic field and is capable of being used for biomagnetic field measurements [6].

There are several ways to induce the magnetization dynamics oscillation. Typically, a coplanar waveguide (CPW) consisted of one signal line and two ground lines is used as a high frequency waveguide. A strong magnetic field induces local magnetization dynamics at around the signal line, which has been widely used to study the magnetization dynamics of magnetic films and sub micro magnetic elements [7-10].

The design of the waveguide is a key factor because it controls magnetic field dependence of magnetization dynamics in ferromagnetic materials. In our previous work, the magnetic field dependence of reflected RF power in Bi doped YIG (Bi-YIG) single crystal thin films was measured using an asymmetrical CPW, and multiple peaks corresponding to the excitation of magnetization dynamics were observed [6]. Those height of peak profile were strongly influenced by the width of each line and the distance between the lines. At a critical condition, the peak height changed drastically. This change could help detecting a small change in magnetic field. However, the study has not completely explained the origin of the multiple peaks.

In this paper, we report magnetization dynamics in a Bi-YIG single crystal thin film induced by a slot line waveguide to understand the origin of the multiple peaks. The slot line waveguide is a representative planar waveguide including two lines with the same width. We systematically changed the width of the lines and the distance between the lines and measured the high frequency power response. The results showed that the interference between local magnetization dynamics oscillation and spin waves generated around each line was a crucial rule for the excitation of magnetization dynamics of the system.

Bi-YIG (111) single crystal thin films with a thickness of about 10 μm were epitaxially grown on gadolinium gallium garnet (GGG) (111) single crystal substrates using liquid phase epitaxy technique. To evaluate the high-frequency response of the samples, Cu slot line waveguides with a thickness of 500 nm were directly fabricated on the Bi-YIG film using a combination of photolithography, Ar ion milling and sputtering techniques. The width and distance between the lines were changed and the length of the sample's signal line(s) and ground lines was fixed at 300 μm (Fig.1 (a)). Since all the waveguides were fabricated on a single Bi-YIG thin film, the magnetic property was constant. When magnetization dynamics was resonantly excited, the input power was absorbed for the use of the precession motion of magnetization, resulting in the significant reduction of the reflected signal strength. Magnetization dynamics was evaluated by measuring the reflected power from the slot line waveguide with the vector network analyzer (Keysight Model N5230A) while sweeping the magnetic field parallel to the sample plane. The input signal strength was set at 0 dBm for all the measurements and the frequency was fixed at a frequency between 6.0 and 9.0 GHz for each measurement. It should be noted that the short length of lines help high-frequency input signals propagating along the waveguide, although the waveguide impedance was different from its measurement setup impedance. The numerical simulations based on micromagnetics were performed using object oriented micromagnetic framework (OOMMF) [11].

Figure 2 (a) shows the reflected high frequency signal strength of various input frequency according to the distance between the lines. Multiple peaks were clearly detected in all conditions. The peaks shifted towards higher magnetic field with the increase of frequency. This indicates that the peaks are associated with magnetic phenomenon. For the samples with the same line width, the number of multiple peaks increased and shifted towards higher magnetic field with the increase of the distance between the lines. Furthermore, one of the peaks shifted higher magnetic field with increasing the width for the samples with the same line's distance as shown in Fig.2 (b).

To understand the experimental results and the origin of the peaks for slot lines waveguides, we carried out the micromagnetics simulation on the RF line's number dependence of magnetization dynamics. The simulation models are shown in Fig. 3 (a). Those names are one-line and two-line models. The dimensions used in the simulation were 200 μm long ($l$), 320 μm wide ($w$), and 10 μm thick (*z*), and one or two lines were arranged. The cell sizes were set at *l*=50 nm, *w*=200 μm and *z*=1 μm. The length of the lines was fixed at 200 μm, and the width and the distance (d) were varied. The Gilbert damping factor α of the Bi-YIG thin film was set at 0.001. A small RF magnetic field (0.5 Oe) was applied perpendicular locally at the position of the lines with the frequency at 8.0 GHz. A DC magnetic field was applied parallel to the length direction to the film. The simulation results are shown in Fig 3 (b). Only one peak associated with FMR is observed for the one-line model. The FMR conditions are determined by the magnetization properties and line width. The broad FMR peak is due to the inhomogeneous demagnetizing field caused by the limitation of simulation size. On the other hand, multiple peaks were identified for the two-line models. When the distance between the lines are identical, the peak shifted toward higher magnetic field with the increase of line widths. The magnetic field where magnetization reaches its peak shifts with the increase of the distance between the lines. Those results faithfully reproduce the experimental results. The difference of the magnetic fields for the appearance of the peaks between the experimental and the simulation results is caused by the difference of the internal magnetic field owing to the limitation of the simulation size. In the experiments, the slot line waveguides with the length of 300 μm were directly fabricated on the Bi-YIG thin film with the substrate size of 11 mm × 11 mm. We simulated the magnetization orientation distribution for x-direction for one-line model as shown in Fig. 3 (c). The DC magnetic field was set at 3.95 kOe, which was the condition for the appearance of the peak of the two-line model in Fig.3 (b). This shows that the line induces local magnetization dynamics. The oscillation of localized magnetization dynamics propagates perpendicular to the line direction via exchange or dipole interaction, and it generates magneto static surface spin waves (MSSW) [12]. This result reveals that the lines also work as spin wave sources. Figure 3 (c) also shows the combined wave of one-line and the 40 µm shifted waves as the green and red lines, respectively. The geometric condition of the combined wave is the same as that of the two-line model. Figure 3 (d) shows the comparison between the combined wave and the spin wave for the two-line model under the same simulation condition. The amplitude of the spin wave for the two-line model is larger than that for the combined wave. This indicates that the local magnetization dynamics oscillation is more amplified in the two-line model.

For the two-line model, a creation of spin waves from the lines propagates toward both sides of the lines, and the spin waves reach the other side of the line. The wavelength of MSSW is described as [13-14]

$$f_{\text{MSSW}} = \sqrt{(f_H + f_M/2)^2 - (f_M/2)^2 \exp(-2kd_0)}, \qquad (1)$$

where $f_H = \gamma H_0$ and $f_M = 4\pi\gamma M_S$ and $d_0$ is film thickness. Here, $\gamma = 2.8$ MHz Oe$^{-1}$ is the electron gyromagnetic ratio, k is the wavenumber of spin wave, $H_0$ is the effective internal field, $M_S = 140$ emu/cc is the saturation magnetization of YIG [12]. The wave length increases with the increase in the applied magnetic field. Fig. 4 shows the experimental results for the magnetic fields at the magnetization peaks to the calculated dispersion curve of the equation (1) for several input frequencies. The peaks appeared when the wave length of the spin wave was almost the same as the distance of the lines. This result indicates that the spin wave generation is potentially relevant with the excitement of magnetization dynamics.

If there is an interaction between the local magnetization dynamics and the spin waves, the condition of the interference should be influenced by the phase difference of those dynamics. The phase difference can be set by the input RF magnetic fields condition of the lines. The simulation results are shown in Fig.5. The peaks shifted with the change of the phase difference. This strongly indicates the existence of the interaction between the local magnetization dynamics and the spin waves.

The possible mechanism of the excitation of magnetization dynamics is as follows. Each local magnetization dynamics under each line generates the spin wave. This spin waves propagate each other and interact with the other side of local magnetization dynamics. When the phase between the local magnetization dynamics and the spin wave coincides each other, the local magnetization dynamics is amplified resonantly. The amplified magnetization dynamics emits the spin wave with large amplitude. The enhanced spin waves propagate and amplify the local magnetization dynamics under the other side of the line. Thus, the local magnetization dynamics under the both side of lines are resonantly excited via the spin waves. This is a typical mutual synchronization phenomenon and have been reported in different systems [15-20]. The reason why we did not detect the FMR peak in our system was that the wave length of spin waves at FMR condition was much longer than the distance between the lines. Thus, the local magnetization dynamics directly influenced each other via the spin waves. We found that the phase difference of the slot line waveguide was about 180° based on numerical electromagnetic analysis. This phase difference effectively suppressed the excitation of the magnetization dynamics for both lines.

Our experimental results show that the magnetization dynamics can be controlled by using multiple sources of local magnetization dynamics. This finding gives the artificial control of the emission condition of highly coherent spin waves without using the FMR, which is attractive for the applications that use the interference of spin waves such as logic gates devices [21-23] and magnonic crystals [24-26].

In conclusion, we studied magnetization dynamics of single Bi-YIG thin films by measuring the high frequency power response induced by the slot lines waveguide. Multiple peaks corresponding to the excitement of magnetization dynamics appeared. We found that the peaks were strongly influenced by the width and the distance of lines. Micromagnetics simulation successfully reproduced the experimental results and revealed that the spin waves generated from the local magnetization dynamics at each line propagated and interfered with the other side of the local magnetization. At a critical condition, this interference resonantly excites the both side of magnetization dynamics owing to the synchronization via spin waves. This finding gives a way for the excitement of the local magnetization dynamics without the FMR condition, which is applicable as a highly coherent spin waves source.

ACKNOWLEDGMENT

We thank GRANOPT Co,Ltd. for their support of Bi-YIG single crystal thin films. This work was supported by JSPS KAKENHI Grant Number JP18K14114.

References

[1] S. I. Kiselev, J. C. Sankey, I. N. Krivorotov, N. C. Emley, R. J. Schoelkopf, R. A. Buhrman, and

D. C. Ralph, Nature **425**, 380 (2003).

[2] W. H. Rippard, M. R. Pufall, S. Kaka, S. E. Russek, and T. J. Silva, Phys. Rev. Lett. **92**, 027201 (2004).

[3] D. Grollier, D. Querlioz, and M. D. Stiles, Proc. of the IEEE **104**, 2024 (2016).

[4] S. Tsunegi, T. Taniguchi, K. Nakajima, S. Miwa, K.Yakushiji, A. Fukushima, S. Yuasa, and H. Kubota, Appl. Phys. Lett. **114**, 164101 (2019).

[5] T. Kanao, H. Suto, K. Mizushima, H. Goto, T. Tanamoto, and T. Nagasawa, Phys. Rev. Appl. **12**, 024052 (2019).

[6] T. Koda, S. Muroga, Y. Endo, IEEE Trans. Magn. **55**, 4002604 (2019).

[7] G. Counil, P. Crozat, T. Devolder, C. Chappert, S. Zoll, and R. Fournel, IEEE Trans. Magn. **42**, 3321 (2006).

[8] I. Neudecker, G. Woltersdorf, B. Heinrich, T. Okuno, G. Gub- biotti, and C. Back, J. Magn. Magn. Mater. **307**, 148 (2006).

[9] C. Bilzer, T. Devolder, P. Crozat, C. Chappert, S. Cardoso, and P. Freitas, J. Appl. Phys. **101**, 074505 (2007).

[10] I. Harward, T. O'Keevan, A. Hutchison, V. Zagorodnii, and Z. Celinski, Rev. Sci. Instr. **82**, 095115 (2011).

[11] OOMMF User's Guide, Version 1.0, M. J. Donahue, D. G. Porter, Interagency Report NISTIR 6376, National Institute of Standards and Technology, Gaithersburg, MD, 1999. http://math.nist.gov/oommf

[12] D. D. Stancil, and A. Prabhakar, *Spin Waves: Theory and Applications* (NewYork, Springer, 2009).

[13] D. D. Stancil, *Theory of MagnetostaticWaves* (NewYork, Springer, 1993)

[14] B. A. Kalinikos, and A. N. Slavin, J. Phys. C: Solid State. Phys. **19**, 7013 (1986).

[15] S. Kaka, M. R. Pufall, W. H. Rippard, T. J. Silva, S. E. Russek, and J. A. Katine, Nature **437**, 389 (2005).

[17] F. B. Mancoff, N. D. Rizzo, B. N. Engel, and S. Tehrani, Nature **437**, 393–395 (2005).

[18] V. E. Demidov, H. Ulrichs, S. V. Gurevich, S. O. Demokritov, V. S. Tiberkevich, A. N. Slavin, A. Zholud and S. Urazhdin, Nat. Commun. **5**, 3179 (2014).

[19] A. Houshang, E. Iacocca, P. Dürrenfeld, S. R. Sani, J. Åkerman and R. K. Dumas, Nat. Nanotechnol. **11**, 280–286 (2016).

[20] A. A. Awad, P. Dürrenfeld, A. Houshang, M. Dvornik, E. Iacocca, R. K. Dumas and J. Åkerman, Nat. Phys. **13**, 292 (2017).

[21] Chumak, A. V., Serga, A. A. & Hillebrands, B., Nature Commun. 5, 4700 (2014).

[22] K. Vogt, F.Y. Fradin, J.E. Pearson, T. Sebastian, S.D. Bader, B. Hillebrands, A. Hoffmann and H. Schultheiss, Nat. Commun. **5**, 3727 (2014).

[23] T. Schneidera, A. A. Serga, B. Leven, and B. Hillebrands, Appl. Phys. Lett. **92**, 022505 (2008).

[24] G Gubbiotti, S Tacchi, M Madami, G Carlotti, A O Adeyeye, and M Kostylev, J. Phys. D **43**, 264003 (2010).

[25] Y. V. Gulyaev, S. A. Nikitov, L. V. Zhivotovskii, A. A. Klimov, P. Tailhades, L. Presmanes, C.

Bonningue, C. S. Tsai, S. L. Vysotskii, and Y. A. Filimonov, JETP Lett. **77**, 567–570 (2003).

[26] M. Krawczyk, and D. Grundler, J. Phys. Cond. Matter **26**, 123202 (2014).

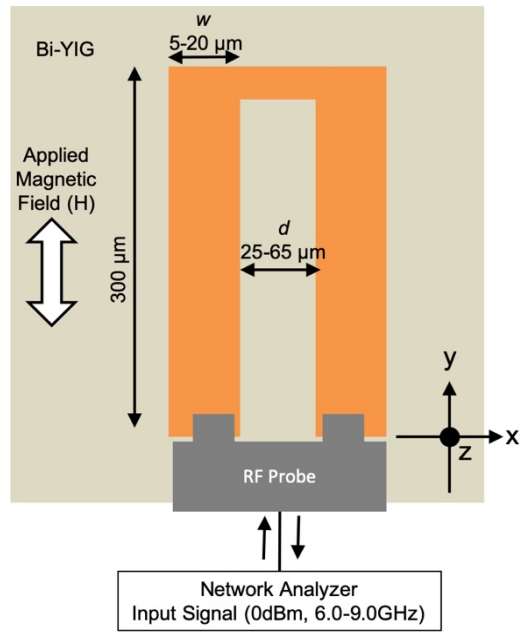

Fig. 1 Schematic view of measurement set up for magnetization dynamics using a slot line waveguide.

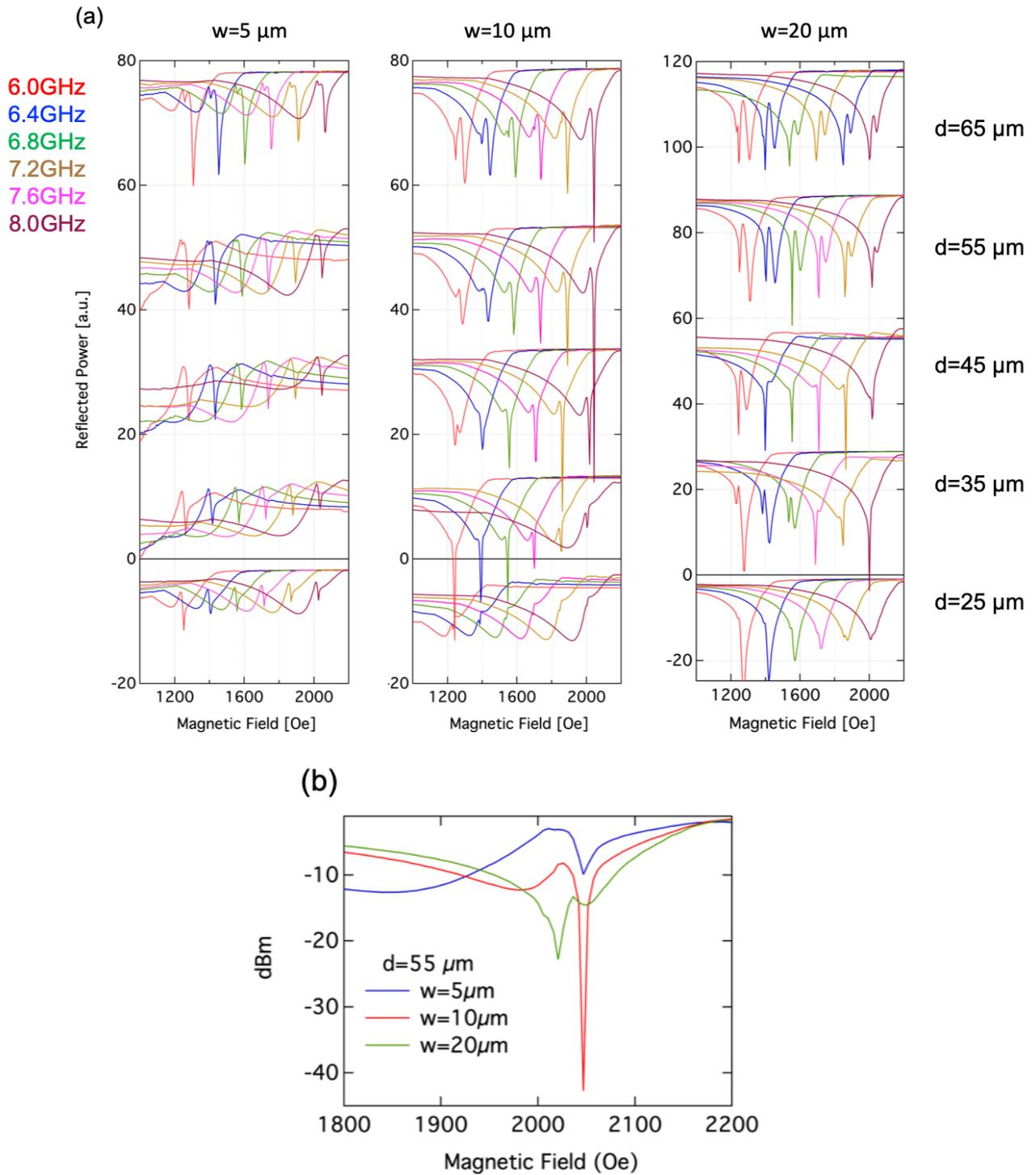

Fig. 2 (a) Applied magnetic field dependence of the power of reflected high frequency wave with various frequency of input power. The distance between lines and the width of the line were systematically changed. (b) Slot line waveguide's width dependence of the reflected power measured at 8.0 GHz.

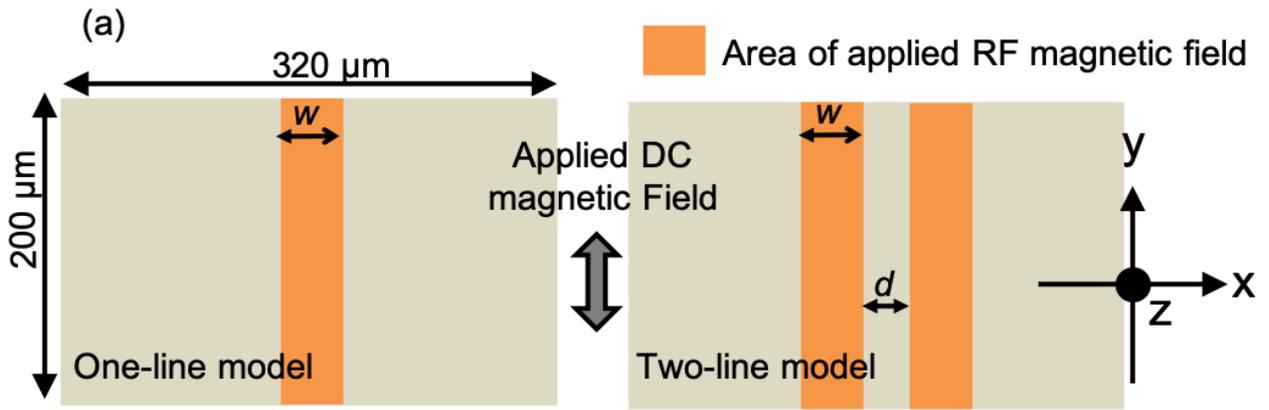

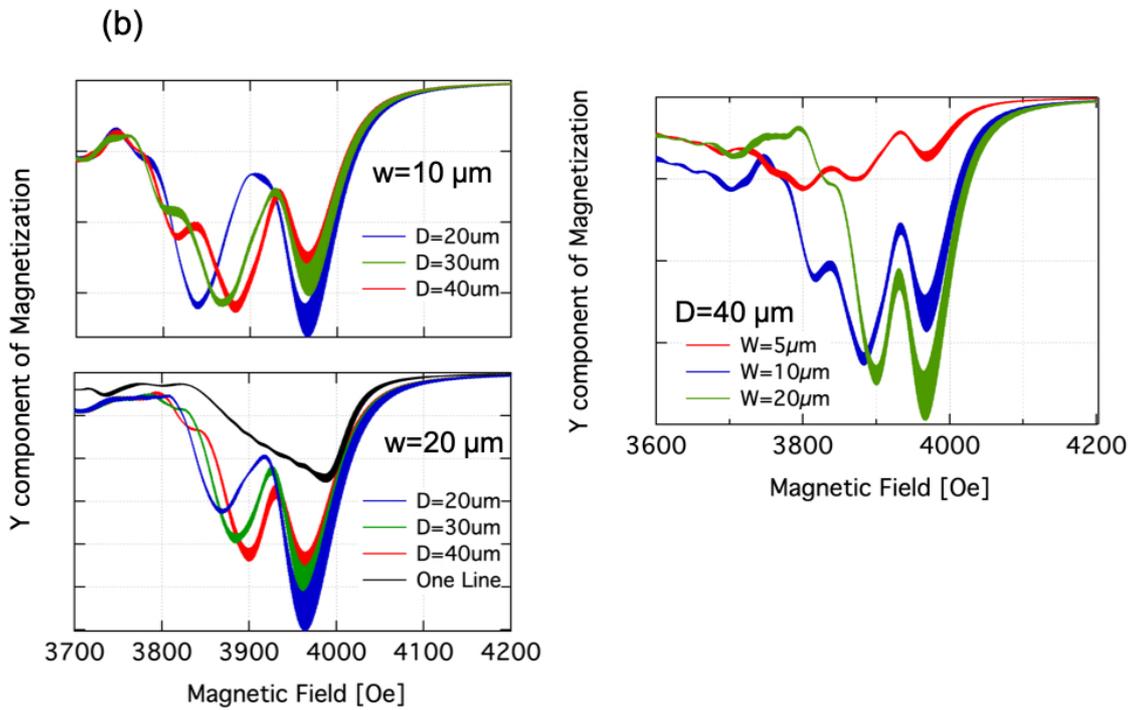

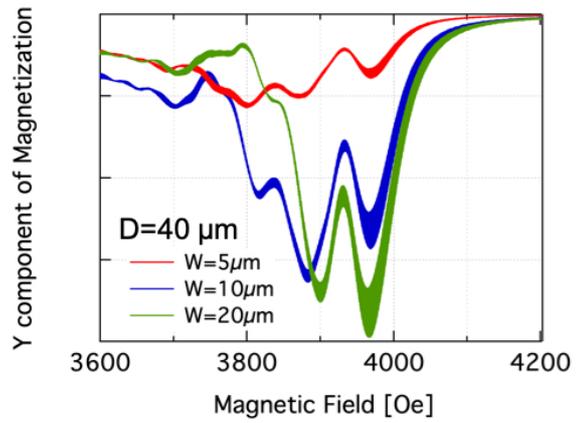

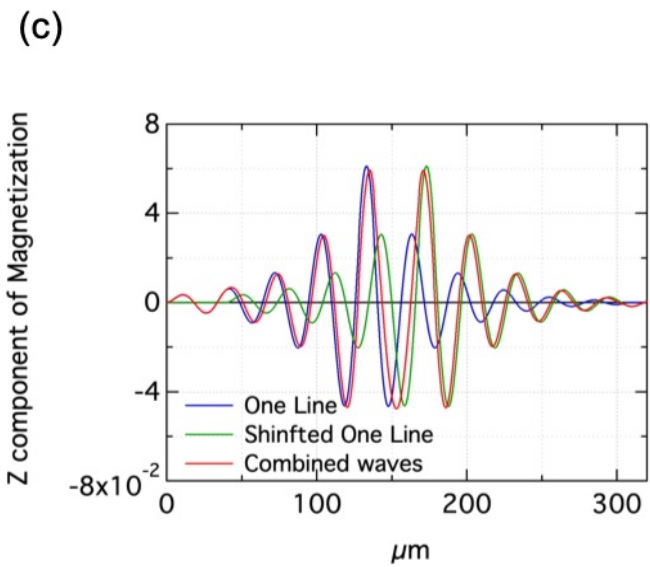

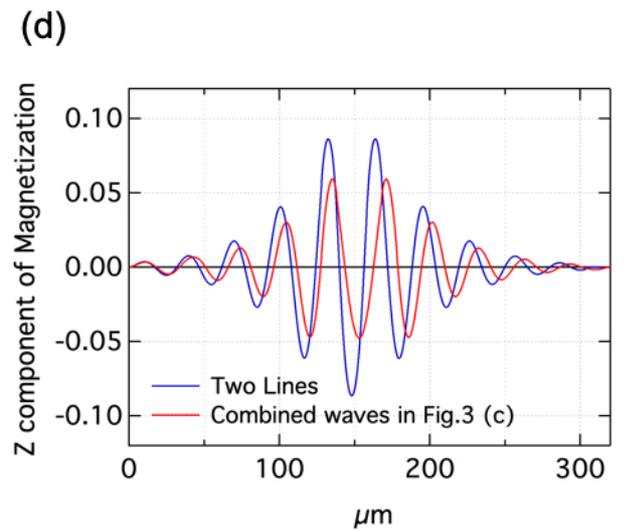

Fig. 3 (a) Schematic view of micromagnetics simulation models. The number of local areas applied RF magnetic field was set at one or two. The dimensions used in the simulation were 200 μm long, 320 μm wide, and 10 μm thick. and one or two lines were arranged. The length of the lines was fixed at 200 μm, and the width (*w*) and the distance (*d*) were varied.

(b) Micromagnetic simulation results for the line's number dependence of magnetization dynamics. W and D stand for the width of lines and the distance between lines, respectively.

(c) Magnetization orientation distribution for x-direction for one-line model at 3.95kOe. Y-axis indicates the z component of normalized magnetization at each position. The green line shows the 40 μm shifted curve of the blue line. The red line shows the combined curve of the blue and green curves.

(d) Comparison for simulation results of the spin waves for two-lines model and the combined curve shown in Fig.3 (c).

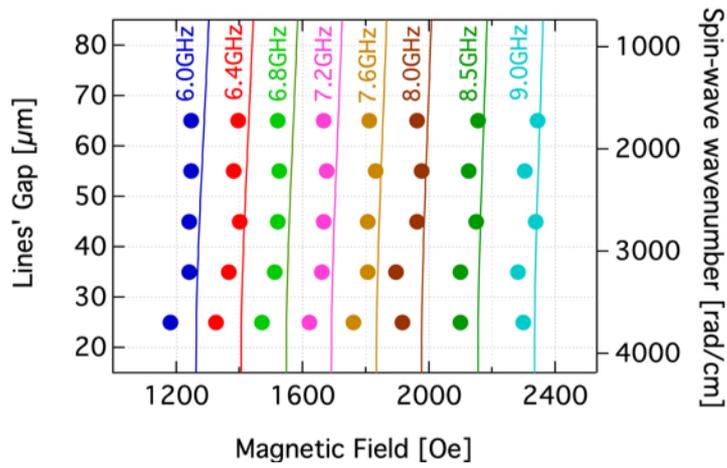

Fig. 4 Lines' distance dependence of magnetic field for the appearance of magnetization dynamics. The solid lines show the calculated dispersion curves using the equation (1).

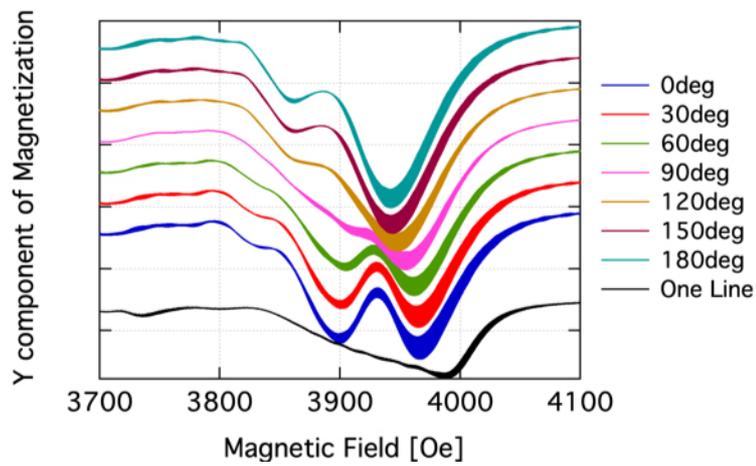

Fig. 5 Magnetization dynamics of the two-line model. The input RF magnetic fields have various range of phase difference.

17